\documentclass[journal]{IEEEtran}

\ifCLASSINFOpdf
\else
  \usepackage[dvips]{graphicx}
\fi
\usepackage{url}

\usepackage{color}
\usepackage{graphicx}
\usepackage{amsfonts}
\usepackage{amssymb}
\usepackage{stfloats}
\usepackage{cite}
\usepackage{psfrag}
\usepackage{subfigure}
\usepackage{amsmath}
\usepackage{array}
\usepackage{algorithm}
\usepackage{algpseudocode}
\usepackage{setspace}
\usepackage[english]{babel}
\usepackage{blindtext}
\usepackage{url}
\usepackage{epstopdf}
\usepackage{verbatim}
\usepackage{color}
\usepackage{amsmath}
\usepackage{bm}
\usepackage{multirow}
\usepackage{booktabs}
\usepackage{makecell}
\usepackage{lineno}

\usepackage{caption}
\newtheorem{Prob}{Problem}

\begin{document}

\title{Quality of Experience Optimization for Real-time XR Video Transmission with Energy Constraints}

\author{Guangjin Pan, Shugong Xu, \IEEEmembership{Fellow, IEEE}, Shunqing Zhang, \IEEEmembership{Senior Member, IEEE},\\ Xiaojing Chen, \IEEEmembership{Member, IEEE}, and Yanzan Sun, \IEEEmembership{Member, IEEE}
\thanks{This work was supported in part by the National Key R\&D Program of China under Grant 2022YFB2902005, the National High Quality Program grant TC220H07D, the National Natural Science Foundation of China (NSFC) under Grant 61871262, 62071284, and 61901251, the National Key R\&D Program of China grants 2022YFB2902000, Foshan Science and Technology Innovation Team Project grant FS0AAKJ919-4402-0060. The corresponding author is Shugong Xu.}
\thanks{G. Pan, S. Xu, S. Zhang, X. Chen, and Y. Sun are with School of Information and Communication Engineering, Shanghai University, Shanghai, 200444, China (e-mail: \{guangjin\_pan, shugong, shunqing, jodiechen, yanzansun\}@shu.edu.cn).}
}

\markboth{IEEE TRANSACTIONS ON VEHICULAR TECHNOLOGY, Vol. X, No. X, March 2023}
{Shell \MakeLowercase{\textit{et al.}}: Bare Demo of IEEEtran.cls for IEEE Journals}
\maketitle

\begin{abstract}
Extended Reality (XR) is an important service in the 5G network and in future 6G networks. In contrast to traditional video on demand services, real-time XR video is transmitted frame-by-frame, requiring low latency and being highly sensitive to network fluctuations. In this paper, we model the quality of experience (QoE) for real-time XR video transmission on a frame-by-frame basis. Based on the proposed QoE model, we formulate an optimization problem that maximizes QoE with constraints on wireless resources and long-term energy consumption. We utilize Lyapunov optimization to transform the original problem into a single-frame optimization problem and then allocate wireless subchannels. We propose an adaptive XR video bitrate algorithm that employs a Long Short Term Memory (LSTM) based Deep Q-Network (DQN) algorithm for video bitrate selection. Through numerical results, we show that our proposed algorithm outperforms the baseline algorithms, with the average QoE improvements of 0.04 to 0.46. \textcolor{black}{Specifically, compared to baseline algorithms, the proposed algorithm reduces average video quality variations by 29\% to 50\% and improves the frame transmission success rate by 5\% to 48\%.}
\end{abstract}


\IEEEpeerreviewmaketitle

\section{Introduction}

\IEEEPARstart{E}{xtended} reality (XR) is one of the crucial applications of future networks. With the rise and maturity of applications such as virtual reality (VR), augmented reality (AR), mixed reality (MR), cloud gaming, and volumetric video \cite{cite:XR2}, people's demand for experience in XR applications is increasing. To meet the growing demand for high-quality XR experiences, real-time video communication technology is essential. This involves providing communication services with high bandwidth and low latency over the network \cite{cite:XR2}. 



However, compared with the chunk-based on-demand video, real-time XR video transmission faces greater challenges. Firstly, due to the high interactivity of XR applications, real-time XR video transmission needs to meet the requirement of low latency, which is often at the frame level \cite{cite:3GPPXR}. Secondly, the frame-by-frame transmission increases the sensitivity of real-time video to network fluctuations \cite{cite:Loki}. When the wireless environment changes, the network conditions may fluctuate. Overestimating network conditions and choosing a higher bitrate for real-time XR video can cause frame loss and stalling. Conversely, underestimating network conditions and selecting a lower video bitrate can decrease video quality. Thirdly, in contrast to the long buffer of on-demand videos (usually over 10 seconds) \cite{cite:NOMA}, in order to ensure the real-time nature of the video, the real-time XR videos typically have a more stringent millisecond-level buffer \cite{cite:Loki}. It results in XR servers not having enough time to acquire buffering information from XR devices for bitrate adaptation decisions. Finally, the network needs to choose a balance between transmission and energy consumption. These challenges prompt us to further research real-time XR video transmission in wireless networks.

To enhance the quality of experience (QoE) in video transmission, prior studies\cite{cite:BOLA,cite:Pensieve,cite:QoE} employ bandwidth prediction, video buffer size, or both to adapt video chunk bitrates. These works do not take into account the role of wireless systems. The authors in \cite{cite:NOMA} use network-assisted adaptive bitrate video streaming (ABS) to improve multi-user capacity region and users’ QoE in non-orthogonal multiple access (NOMA) systems. In \cite{cite:UAV}, the authors study an unmanned aerial vehicle (UAV) enabled video streaming system with dynamic adaptive streaming over HTTP (DASH). However, as mentioned above, chunk-based adaptive bitrate algorithms \cite{cite:BOLA,cite:Pensieve,cite:NOMA,cite:UAV,cite:QoE} cannot be directly applied to real-time XR video applications. To optimize real-time video transmission, Google proposed the Google Congestion Control (GCC) congestion control algorithm for web real-time communication (WebRTC) system \cite{cite:GCC}. In \cite{cite:realtimeUAV2}, the authors aim to optimize live video streaming from UAVs by optimizing the bitrate, motion control, and power control. However, they do not consider the wireless resource allocation for real-time XR video transmission.

We consider a real-time XR video transmission system with a single user\footnote{For single-user scenarios, wireless resource competition is simplified to the constraint on the maximum number of subchannels. It assumes all resources are dedicated to a single user, without considering competition from others.}. The contributions of our work are as follows,
\begin{itemize}
  \item We consider the transmission of XR video in the Orthogonal Frequency Division Multiple Access (OFDMA) system. We propose a QoE model which is suitable for real-time transmission systems that transmit frame-by-frame, including AR, VR, etc. We model the problem as a bitrate selection and subchannel allocation problem for XR video transmission, subject to long-term energy constraints. The optimization objective is to improve the QoE of real-time XR video in the wireless system.
  \item We apply the Lyapunov method to transform the long-term optimization problem into a single-frame optimization problem. The single-frame problem is decomposed into two subproblems, i.e., the subchannel allocation subproblem and the bitrate selection subproblem, which are solved separately. In particular, we propose a Long Short Term Memory (LSTM) based Deep Q-Network (DQN) algorithm to solve the bitrate selection subproblem.
  \item We verify the proposed adaptive real-time XR video streaming method in both a simulated environment and a real system environment. \textcolor{black}{The simulation experiment shows that compared to the baseline algorithm, the proposed LSTM-DQN algorithm, at the cost of average video bitrate reduction, decreases the average video quality variations by 29\% to 50\% and improves the frame transmission success rate by 5\% to 48\%, resulting in an enhancement of average QoE by 0.04 to 0.46.}
\end{itemize}

\section{System model} \label{sec2}

\begin{table}[t]
    \caption{SUMMARY OF MAIN NOTATIONS}
\centering
\begin{tabular}{|c|c|} \hline  \textbf{Notation}& \textbf{Description}\\ \hline
$F$
& Frame rate of XR videos\\ \hline 
 $r_f^{sc}$
& Data rate of each subchannel\\ \hline 
 $B_w$
& Bandwidth of each subchannel\\ \hline 
$h_f$& Channel coefficient\\ \hline
 $\textcolor{black}{p} / N_0$& Transmission power / Noise power\\ \hline 
$D_f$&  Transmission delay of frame $f$\\ \hline 
$\Delta t$& Video frame interval\\ \hline  
 $N_f^{sc}$&Number of allocated subchannel\\ \hline 
 $E_f$ & Transmission energy consumption\\ \hline 
 $v_f$& Data size of frame $f$\\ \hline 
 $x_f$& Frame transmission status\\ \hline 
 $\mathcal{V}$ & Set of available bitrates\\ \hline 
 $\mathcal{G}$ & Set of video frame data sizes\\ \hline
 $o$&Size of the observation window\\ \hline
 $q_f$&QoE for real-time XR video transmission\\ \hline
\end{tabular}
\label{table:Notation}
\vspace{-3mm}
\end{table}

\begin{figure}[tb]
\centering 
\includegraphics[height=2.0in,width=2.4in]{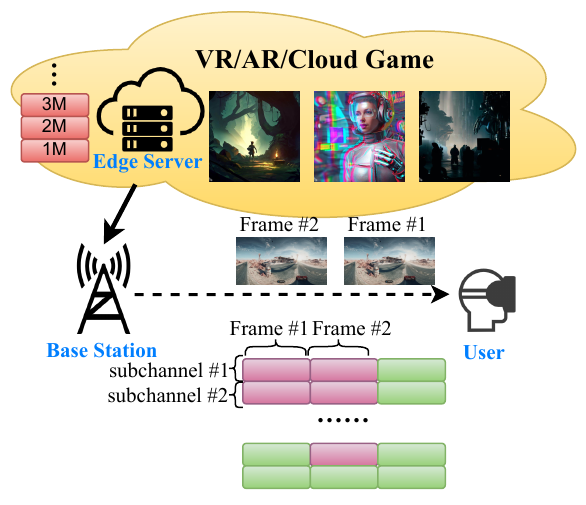} 
\caption{System architecture for adaptive real-time XR video streaming.}
\label{fig:architecture} 
\vspace{-5mm}
\end{figure}

The network architecture is shown in Fig. \ref{fig:architecture}, where the edge server renders XR videos in real-time and transmits them frame-by-frame to the XR user. The base station (BS) allocates subchannels for the transmission of each frame through subchannel allocation. Due to network fluctuations and energy constraints, edge servers need to select the bitrate for each video frame to optimize user QoE. Denoting $F$ as the frame rate of the XR video, we assume that subchannels are scheduled at the granularity of video frames. Each scheduling period has a length of $\Delta t=\frac{1}{F}$, which is also the video frame interval. Table \ref{table:Notation}
summarizes the main notations of this paper.

\subsection{Transmission Model}

During the transmission period of the $f^{th}$ frame, we assume the maximum number of available wireless subchannels is $N^{sc}_{f,max}$. Let $r_f^{sc}$ denote the data rate that each subchannel can achieve. According to \cite{cite:rate}, $r_f^{sc}$ can be expressed as
\begin{eqnarray}
r_f^{sc}&=&B_w \cdot \log_2\left(1+\frac{\textcolor{black}{p}\cdot h^2_f}{N_0\cdot B_w}\right), \label{equ:rate}
\end{eqnarray}
where $B_w$ represents the bandwidth of each subchannel, $\textcolor{black}{p}$ and $N_0$ are the normalized transmission power and noise power, respectively. $h_f$ is the channel coefficient during the transmission of the $f^{th}$ frame of the video, which we assume remains constant during this period of transmission \cite{cite:Hconstant}.

Assuming that the $f^{th}$ frame requires $N_f^{sc}$ subchannels for transmission, the delay $D_f$ of frame $f$ is given by
\begin{eqnarray}
D_f&=&\frac{v_f}{r_f^{sc}N_f^{sc}}, \label{equ:delay}
\end{eqnarray}
where $v_f$ is the data size of the frame $f$.

In video frame transmission, a delay threshold $\tau$ is imposed to define the maximum delay allowed between the arrival and successful transmission of frames. \textcolor{black}{ A transmission failure occurs if the transmission delay exceeds $\tau$. Considering the limited buffer length in real-time XR devices, we set $\tau=\Delta t$ which means that each $f^{th}$ video frame must be fully transmitted before the arrival of the next $(f+1)^{th}$ frame.}

Therefore, we define an indicator function $x_f$ to represent the frame transmission status (successful or failed), i.e.,
\begin{eqnarray}
x_f&=& \left\{
{\begin{aligned}
0 ,  \qquad \mathop{\textrm{if}} D_f \leq \Delta t,\\
1 ,   \qquad \ \ \mathop{\textrm{otherwise}}.
\end{aligned}}  \right. \label{equ:frame_success}
\end{eqnarray}

For the transmission of each frame, the energy consumption $E_f$ is related to the number of allocated subchannels and transmission time. For the frame $f$, $E_f$ can be given by \cite{cite:energy}
\begin{eqnarray}
E_f&=&\textcolor{black}{p} N^{sc}_f \cdot \textrm{min}\{D_f,\Delta t\}, \label{equ:energy}
\end{eqnarray}
where $\textrm{min}\{D_f,\Delta t\}$ means that the transmission time of this frame does not exceed $\Delta t$ at maximum.
\color{black}

\subsection{Real-time XR Video Streaming and QoE Model} \label{sec:qoemodel}

Let $\mathcal{V}$ denote the set of available bitrates with $M$ bitrate levels, where $\mathcal{V}=\{V_1,V_2,\ldots,V_M\}$ and $V_1 \le V_2 \le \ldots \le V_M$. For the $f^{th}$ frame, the frame data size corresponding to the bitrate is $v_f\in \mathcal{G} = \{\frac{V_1}{F},\frac{V_2}{F},\ldots,\frac{V_M}{F}\}$ \footnote{In practical systems, the size of video frames strongly depends on the compression techniques. For example, the size of I-frames is usually larger than that of P-frames and B-frames. To simplify the problem, we assume that the impact of video encoding on frame size is neglected in this paper.}. 


The chunk-based adaptive video streaming models QoE using bitrate, variations in bitrate, and rebuffering time \cite{cite:QoE}. \textcolor{black}{However, real-time XR video transmission sends only one frame per transmission and has a small buffer length, making it unable to utilize buffer length as a metric for measuring QoE. Referring to the QoE model of chunk-based video streaming, the proposed novel QoE model for adaptive real-time XR video streaming is constructed based on the following aspects,}
\begin{itemize}
  \item{\textit{Video Quality}}. We define the QoE for video quality as the average bitrate of each frame, e.g., ${v_f}$.
  \item{\textit{Quality variations.}} 
  Inspired by \cite{cite:Loki}, Quality variations use the differences in bitrate adjustments as a component for measuring QoE to represent the stability of video bitrate. It is given by $|v_f-v_{f-1}|$.
  \item{\textit{Frame transmission results}.} Frame transmission failures have a significant impact on the user experience. \textcolor{black}{We utilize frame transmission results instead of the buffer size as a metric to measure the quality of transmission.}
\end{itemize}

Define the weighted sum of the above three components as the QoE $q_f$ of real-time XR video. Taking into account that users have different preferences, the QoE can be written as
\begin{eqnarray}
q_f =(1-x_f)({v_f}-\mu_1 |v_f-v_{f-1}|)-\mu_2 x_f, \label{equ:qoe}
\end{eqnarray}
\textcolor{black}{
where $\mu_1, \mu_2 \in [0,+\infty)$ are weight parameters, which depend on various factors, including the differences in XR video content, codec selection, and the diverse personalized requirements of XR devices.} By defining the QoE, we aim for higher more stable bitrates for XR videos, as well as a higher success rate in video frame transmission.


We define a bitrate selection function $\mathcal{F}(\cdot)$ to select an appropriate video bitrate to maximize the QoE of XR video transmission. However, due to the inability of the edge server to obtain real-time BS status information, $\mathcal{F}(\cdot)$ cannot be expressed in a closed-form using the current network information. To address this issue, we propose to determine the inherent relationship between bitrate selection in high-level and communication modes in low-level based on historical bitrate and frame transmission status to guide the current real-time XR video bitrate selection. Let $\bm{v}_{f-1}=[v_{f-1},v_{f-2},\cdots,v_{f-o}]$ and $\bm{x}_{f-1}=[x_{f-1},x_{f-2},\cdots,x_{f-o}]$ denote the historical bitrate and frame transmission status, respectively. $o$ is the size of the observation window, and is determined based on whether the historical information contains sufficient information to assist the agent in selecting the appropriate video bitrate. The adaptive real-time XR video streaming algorithm can be represented as
\begin{eqnarray}
v_f&=&\mathcal{F}(\bm{v}_{f-1},\bm{x}_{f-1}) \in \mathcal{G}. \label{equ:abr}
\end{eqnarray}
$\mathcal{F}(\cdot)$ is a function to discover potential patterns in bandwidth fluctuations and scheduling mechanisms and select the appropriate video bitrate to improve user QoE. The historical bitrates and frame transmission status can be obtained and cached on the edge server after each frame transmission, enabling the selection of the appropriate bitrate for subsequent frames.
\color{black}




\subsection{Problem Formulation}
This paper aims to optimize the average QoE by accounting for energy consumption and wireless resource constraints.
\begin{Prob}
[Original Problem] The original problem can be formulated as the following optimization problem,
\begin{eqnarray}
\underset{\mathcal{F}(\cdot),N_f^{sc}}{\textrm{maximize}}  &&  \underset{N_{frame}\rightarrow \infty}{\textrm{lim}}  \frac{1}{N_{frame}} \sum_{f=1}^{N_{frame}} {q}_f, \label{eqm:ori_obj}\\
\textrm{subject to} 
                && \underset{N_{frame}\rightarrow \infty}{\textrm{lim}} \frac{1}{N_{frame}} \sum_{f=1}^{N_{frame}}  E_f \leq e,\\
                && \eqref{equ:rate}-\eqref{equ:abr}, \nonumber
                \\
                &&N_f^{sc} \le N^{sc}_{f,max}, \label{eqn:cons_subchannel} \\
                 &&  v_f\in \mathcal{G}, \label{eqn:cons_bitrate}
\end{eqnarray}
\end{Prob}
where $e$ is the long-term average energy constraint.


\color{black}
The problem is in general difficult to solve. Firstly, the optimization objective is the long-term average QoE for XR video transmission. Therefore, the computational complexity of the problem is intricately associated with the number of frames $N_{frame}$. Considering that $N_{frame}$ tends towards positive infinity, the complexity of the solution also approaches infinity, resulting in the difficulty of solving the original problem. Secondly, due to the inability of the video server to acquire BS status information in real-time, the $\mathcal{F}(\cdot)$ cannot be expressed as a closed-form mathematical model, making it difficult to solve for the optimal $\mathcal{F}(\cdot)$. Finally, the dynamic nature of the wireless channel, including fluctuations and variations in the number of subchannels, further exacerbates the difficulty of problem solving. 
\color{black}


\section{Lyapunov-guided DQN algorithm}\label{sec3}

In this section, we aim to solve the optimization problem mentioned above. We propose a Lyapunov-guided LSTM-DQN algorithm to solve this problem.

\subsection{Problem Transformation via Lyapunov Optimization}\label{sec3-1}

To address the difficulty in solving the problem caused by long-term energy constraints, we introduce a virtual queue $Q_f$ and transform it into a queue stability constraint. The update of the virtual queue can be represented as
\begin{eqnarray}
Q_{f+1}=Q_f+E_f-e.  \label{equ:VirtualQueue}
\end{eqnarray}

The Lyapunov function $V_{f}$ is subsequently defined to measure queue congestion by employing a quadratic function \cite{cite:lyapunov},
\begin{eqnarray}
V_{f+1}=\frac{1}{2}Q_{f+1}^2 \leq \frac{1}{2}Q_f^2+ Q_f(E_f-e)+\frac{1}{2}B,  \label{equ:Lyapunovfunction}
\end{eqnarray}
\color{black}
where $B={E_f^{max}}^2+e^2$, and $E_f^{max}=\textrm{max}_f E_f$ is the maximum energy consumption of one frame transmission. Let $\triangle V_f= V_{f+1}- V_f$ denote the Lyapunov drift, which represents the expected change of the Lyapunov function from the current video frame transmission to the next video frame transmission. Combined with \eqref{equ:Lyapunovfunction}, it can be given by
\begin{eqnarray}
\triangle V_f & \leq & \frac{1}{2}Q_f^2+ Q_f(E_f-e)+\frac{1}{2}B - \frac{1}{2}Q_f^2 \nonumber \\ 
&=& Q_f(E_f-e)+\frac{1}{2}B.\label{equ:Lyapunovdrift}
\end{eqnarray}

According to \cite{cite:lyapunov}, minimizing the Lyapunov drift function for every frame is an effective way to preserve queue stability. To achieve this, we need to optimize both the objective function and queue stability simultaneously. The single-frame optimization problem is to minimize the Lyapunov drift-plus-penalty, which is $-\beta {q}_f+\Delta V_f$. $\beta$ is the weight parameter to make a tradeoff between the minimization of the objective function and the stability of the queue. The upper bound of the Lyapunov drift-plus-penalty can be expressed as $-\beta {q}_f+\Delta V_f \leq -\beta {q}_f + Q_fE_f-eQ_f+\frac{1}{2}B.$ Considering that $B$ and $e$ are fixed constants and that $Q_f$ is determined before each frame transmission, they can be ignored. Therefore, the single-frame optimization problem can be expressed as
\begin{Prob}
[Single-frame Optimization Problem] Based on the Lyapunov optimization theory, the original problem can be converted to
\begin{eqnarray}
\underset{\mathcal{F}(\cdot),N_f^{sc}}{\textrm{Minimize}}  &&  \!\!  -\beta {q}_f + Q_fE_f, \\
\textrm{subject to} 
                &&  \eqref{equ:rate}-\eqref{equ:abr},\eqref{eqn:cons_subchannel},\eqref{eqn:cons_bitrate},\nonumber 
\end{eqnarray}
\end{Prob}
To solve the problem, we assume that the bitrate is given and solve the subchannel allocation problem. Then, we propose an LSTM-DQN method to solve the bitrate adaptation problem.
\color{black}


\subsection{Solution for Channel Allocation Subproblem} \label{Sec:ChannelAllocation}
Despite simplifying the problem using Lyapunov optimization, solving the above-mentioned problem remains challenging. Assuming that the function $\mathcal{F}(v_{f-1},{x_f-1})$ is known, we can solve the channel allocation subproblem first.

\begin{Prob}
[Channel Allocation Subproblem] The channel allocation subproblem can be formulated as
\begin{eqnarray}
\underset{N_f^{sc}}{\textrm{Minimize}}  &&  \!\!  -\beta {q}_f + Q_fE_f, \\
\textrm{subject to} 
                &&  \eqref{equ:rate}-\eqref{equ:abr}, \eqref{eqn:cons_subchannel}. \nonumber 
\end{eqnarray}
\end{Prob}
The result of this mixed programming problem can be obtained by solving two cases, one with $x_f=1$ and the other with $x_f=0$. The result of these two problems can be given by
\begin{eqnarray}
N_f^{sc}&=& \left\{
{\begin{aligned}
0 ,  \qquad \quad \qquad  N^{sc}_{f,max} < \frac{v_f}{\Delta t r_f^{sc}} \qquad \qquad \quad   \ ,\\
\frac{v_f}{\Delta t r_f^{sc}} , \   \mu_2 \ge \mu_1 |v_f-v_{f-1}|-{v_f}+\frac{Q_f\textcolor{black}{p}v_f}{\beta\Delta t r_f^{sc}}, \nonumber\\
0 , \qquad \mu_2 < \mu_1 |v_f-v_{f-1}|-{v_f}+\frac{Q_f\textcolor{black}{p}v_f}{\beta\Delta t r_f^{sc}}.
\end{aligned}}  \right.
\end{eqnarray}

The result shows that when the maximum number of subchannels available does not meet the delay requirements (i.e., $N^{sc}_{f,max} < \frac{v_f}{\Delta t r_f^{sc}}$), the optimal solution is to drop the frame entirely. This is because a video frame can only be decoded and played once it has been fully transmitted. Insufficient wireless resources allocated to video transmission result in a waste of resources and increased energy consumption, without enhancing QoE. Additionally, if the assigned number of subchannels is sufficient but leads to increased energy consumption exceeding the QoE improvement from successful frame transmission, frames are also discarded (i.e., $\mu_2 < \mu_1 |v_f-v_{f-1}|-{v_f}+\frac{Q_f\textcolor{black}{p}v_f}{\beta\Delta t r_f^{sc}}$). Therefore, discarding a video frame means not allocating any subchannels for transmitting the data of this frame. Otherwise, wireless resources are allocated to ensure frame transmission, and the strategy is to allocate the minimum number of subchannels for successful transmission and minimal energy consumption.


\subsection{Solution for Adaptive Video Streaming}
To optimize the average QoE, we define the adaptive XR video streaming problem as an infinite-horizon Markov decision process (MDP). The MDP problem has three elements, i.e., the state set $\mathcal{S}$, action set $\mathcal{A}$, and reward $\mathcal{R}$. The goal of reinforcement learning is to select the appropriate actions based on observing the state of the environment to maximize the reward. The details of the MDP are defined as follows, 
\subsubsection{\textbf{State}} The state, which is denoted by $\bm{s}_f=\{\bm{v}_{f-1},\bm{x}_{f-1}\} \in \mathcal{S}$, consists of the historical video frames' bitrate and transmission success indicator.
\subsubsection{\textbf{Action}} The action $a_f \in \mathcal{A}$ is the selected bitrate, and the action space is the same as the available bitrate set $\mathcal{A}=\mathcal{V}$. The frame data size $v_f$ is calculated as the quotient of $a_f$ and frame rate $F$, i.e., $v_f=\frac{a_f}{F}$. Therefore, in this problem, the action represents the data size of each frame.
\subsubsection{\textbf{Reward}} According to the original problem, we formulate the reward function as
\begin{eqnarray}\label{equ: reward} 
r(\bm{s}_f,\bm{a}_f)=q_f.
\end{eqnarray}

In each episode, we use the action network for bitrate selection and then perform channel allocation using the method described in Sec. \ref{Sec:ChannelAllocation}. Based on the result of the channel allocation, we can calculate the reward. We store experiences from the agent, i.e., $m_f=\{\bm{s}_f,\bm{a}_f,r(\bm{s}_f,\bm{a}_f),\bm{s}_{f+1}\}$ to the replay memory. Then, we use the bootstrapped random update method to train the Q-network. Specifically, we sample a mini-batch data $\{m_1, m_2 \cdots m_S\}$ from the replay memory and use the loss function $L(\theta_g)$ to update the Q-network, where
\begin{eqnarray}
L(\theta_g)=\frac{1}{S} \sum_{i=1}^S [y_i(\theta_g')-Q(\bm{s}_i,\bm{a}_i;\theta_g)]^2.
\end{eqnarray}
$S$ is the number of samples, $\theta_g$ is the parameter of the Q-network, $\theta_g'$ is the parameter of a fixed target Q-network, and 
\begin{eqnarray}\label{equ:target}
y(\theta_g')=r(\bm{s}_i,\bm{a}_i)+ \! \! \gamma \mathop{\textrm{max}}_{\bm{a}_{i+1}} \! Q(\bm{s}_{i+1},\bm{a}_{i+1};{\theta}_g').
\end{eqnarray}
$\gamma$ is the discount factor.

To combine historical information, we propose an LSTM-based DQN network for problem solving. The model consists of a cascaded structure with 2 layers of LSTM \cite{cite:LSTM} and 2 layers of fully connected (FC) \cite{cite:realtimeUAV2} layers. In our proposed model, the LSTM has 64 hidden units, the output size of the first FC layer is 512, and the output of the second FC layer is $|\mathcal{V}|$. Compared to FC and Convolutional Neural Network (CNN), the LSTM can capture temporal dependencies, making it suitable for extracting features from a given historical sequence \cite{cite:LSTM}. The proposed LSTM-DQN network can utilize historical sequence information to assist in selecting the appropriate video bitrate.


\section{Numerical Results}
In this section, we provide some numerical results to demonstrate the advantages of the proposed DQN-based adaptive video streaming algorithm. In our simulation setting, the set of available bitrates is $\mathcal{V}=\{1, \ 2 ,\cdots,\ 20\}$ Mbps, and the frame rate is 60FPS. The frame sizes are generated according to the 3GPP standards \cite{cite:3GPPXR}. The bandwidth of each subchannel is 180 KHz, and the transmission power of each subchannel is 200 mW. We set $\mu_1=0.1$, $\mu_2=1$, and $\beta=1$. To compute the QoE numerically, we first obtain the results of subchannel allocation and bitrate selection based on the algorithms described in Sec. III-B and Sec. III-C. Then, we utilize \eqref{equ:qoe} to calculate the QoE for each frame of the XR video. To obtain the average QoE for the entire simulation, we conducted simulations for 10,000 frames and calculated the average QoE by summing up the QoE values of all these frames and dividing them by the total number of frames.


We compare the proposed Lyapunov-guided LSTM-DQN scheme to the following baseline systems. 1). Baseline 1 adopts a simple adaptive adjustment strategy. When the previous frame is successfully transmitted, this strategy increases the bitrate of the $f^{th}$ frame. Otherwise, it decreases the bitrate. 2). Baseline 2 \cite{cite:realtimeUAV2} uses the DQN algorithm based on an FC neural network. It adapts the bitrate based on the previous frame's video bitrate and video QoE. 3). Baseline 3 \cite{cite:Pensieve} uses the Asynchronous Advantage Actor-Critic (A3C) algorithm to select the bitrate. It takes into account the historical video bitrates and frame transmission status in the decision-making process. 4). Baseline 4 \cite{cite:greedy} is a Greedy algorithm with perfect prior knowledge, assuming that the agent has prior knowledge of the network environment, including the maximum number of channels, subchannels, and energy queues. It uses a traversal method to select the bitrate that maximizes the current QoE, and the result obtained is optimal for a single frame.

\begin{figure}[tb]
    \centering
    \includegraphics[height=1.5in,width=2.0in]{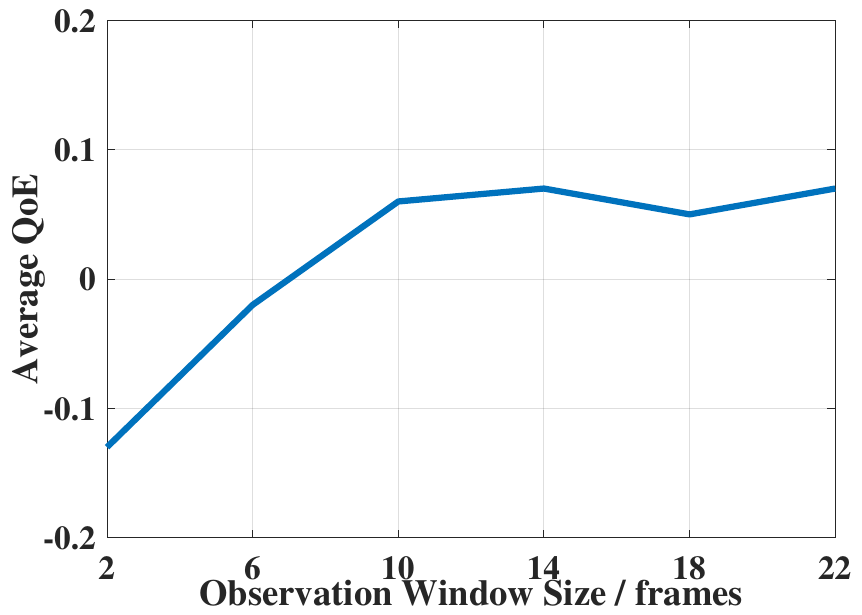}
    \caption{Comparison of the average QoE under different observation window sizes.}
    \vspace{-3mm}
    \label{result_window}
\end{figure}

\subsection{Selection of Observation Window Size}

We explore the impact of different observation window sizes on algorithm performance. For the simulation environment, we use 10 subchannels with an energy constraint of $e=0.03$ J. The user's signal-to-noise ratio (SNR) is set at 20 dB. In Fig. \ref{result_window}, when the observation window size is less than 10, the average QoE increases as the observation window size increases. This implies that a larger observation window provides ample historical data, allowing the model to accurately identify communication patterns and trends. In contrast, for observation window sizes exceeding 10, the average QoE basically remains stable. Considering the computational complexity, we set an observation window size of $o=10$ frames for comparison in the subsequent experiments.
\color{black}


\subsection{Performance Comparison}

\begin{figure}[tb]
\centering 
\includegraphics[height=2.3in,width=2.8in]{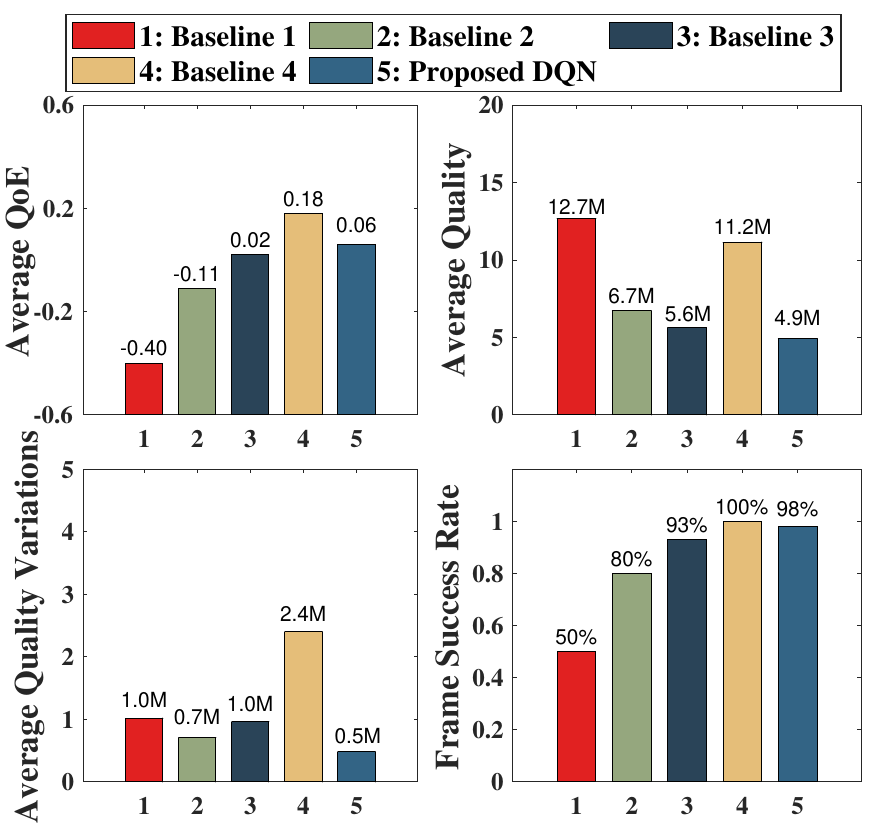} 
\caption{\textcolor{black}{Comparison of average QoE, average quality, average quality variation, and frame transmission success rate between different algorithms in a simulation environment.}}
\label{fig:Simresult} 
\vspace{-4mm}
\end{figure}

\begin{figure}[tb]
\centering 
\includegraphics[height=3.0in,width=2.8in]{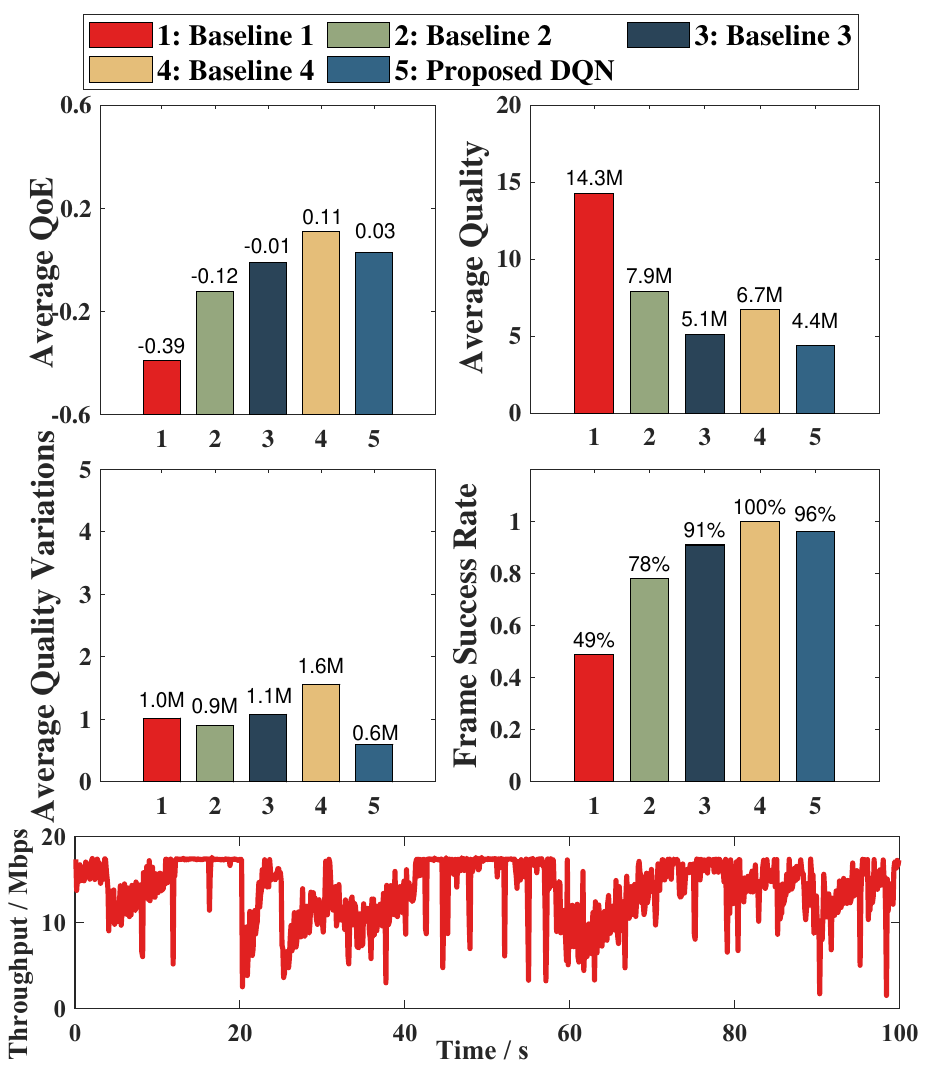} 
\caption{\textcolor{black}{Comparison of average QoE, average quality, average quality variation, and frame transmission success rate between different algorithms in the real system environment. The data traces are collected from the OAI system.}}
\label{fig:OAIresult} 
\vspace{-4mm}
\end{figure}

We conduct performance comparisons in both a simulated and a real system environment. In Fig. \ref{fig:Simresult}, we plotted the average QoE and the components that make up QoE, including average quality, average quality variation, and frame transmission success rate, which are explained in Sec. \ref{sec:qoemodel}. Regarding the average quality, it represents the average bitrate of each video frame. Similarly, the average quality variation measures the difference in bitrate between each video frame and its previous frame, and the frame success rate represents the proportion of successfully transmitted video frames out of the total number of video frames. As demonstrated in Fig. \ref{fig:Simresult}, the average QoE of our proposed DQN algorithm has improved by 0.46, 0.17, and 0.04, when compared to Baselines 1, 2, and 3, respectively, and it is only 0.12 lower than Baseline 4 (Greedy algorithm utilizing prior information). In addition, the average quality of the proposed DQN algorithm is 4.9 Mbps, the average quality variation is 0.5 Mbps, and the frame transmission success rate is 98\%. \textcolor{black}{The analysis of the components of QoE indicates that despite sacrificing average video quality, the proposed algorithm demonstrates superior bitrate stability (quality variations) and frame success rate, resulting in an outstanding QoE performance.}

For the real system environment, the data trace is obtained from the Openairinterface (OAI) system \cite{cite:OAIenb}. We collect data traces of resource blocks and rates, as described in \cite{cite:bandwidth}, and use these traces for algorithm testing. We assess the performance of algorithms using the collected data traces. Fig. \ref{fig:OAIresult} displays the average QoE and the components of QoE, as well as the throughput for the collected data. In the real system environment, the proposed adaptive video streaming scheme achieves an average QoE of 0.03. Compared to Baselines 1-3, the proposed DQN algorithm achieves 0.42, 0.15, and 0.04 average QoE improvements, respectively. In addition, the proposed DQN algorithm has an average quality of 4.4 Mbps, an average quality variation of 0.6 Mbps, and achieves a frame transmission success rate of 96\%. In the actual data trace, we can obtain similar conclusions to those in the simulation environment, which proves the effectiveness of the algorithm.

\subsection{QoE Analysis}

\begin{figure}[tb]
\centering 
\includegraphics[height=1.9in,width=2.4in]{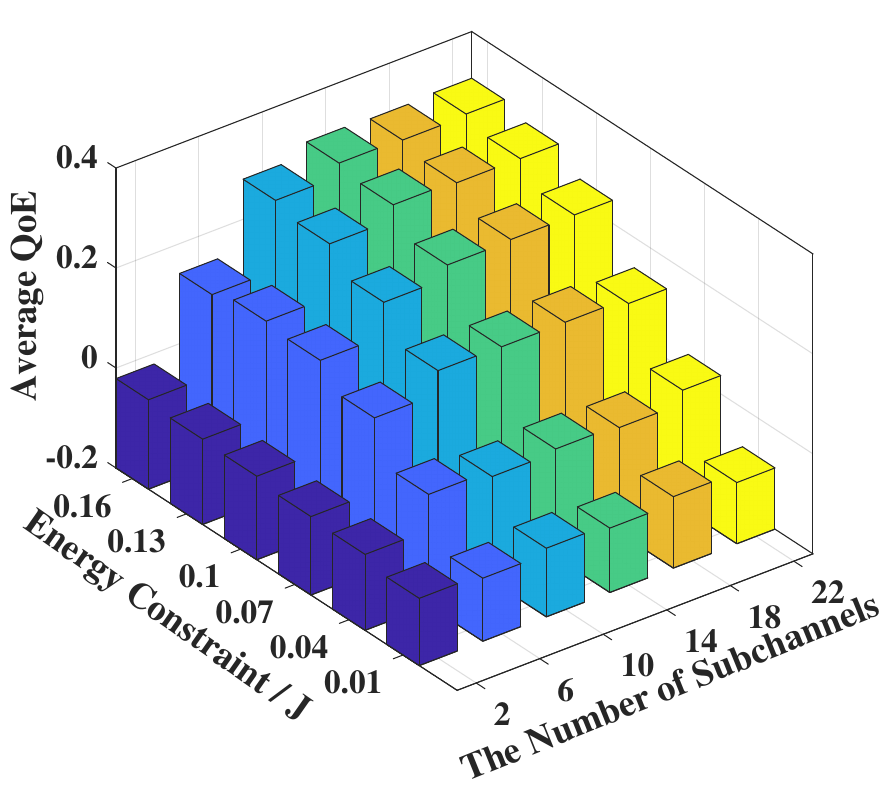} 
\caption{\textcolor{black}{Average QoE of our proposed algorithm under different energy constraints and subchannel constraints.}}
\label{fig:QoESurface} 
\vspace{-3mm}
\end{figure}

We explore the optimal QoE changes of real-time XR video under average energy consumption and wireless resource constraints. As shown in Fig. \ref{fig:QoESurface}, both average energy consumption and wireless resource limitations affect the user's QoE. When wireless resources are sufficient, average energy consumption is the key limiting factor for QoE. Conversely, when average energy consumption is sufficient, subchannel numbers become the limiting factor. \textcolor{black}{When the average energy constraint exceeds 0.1 J and the number of subchannels exceeds 10, the wireless resources can support maximum bitrate transmission, resulting in no further improvement in QoE.}
\color{black}

\section{Conclusion}


This paper consider the problem of adaptive bitrate streaming for wireless transmission of real-time XR video. We propose a QoE model for XR applications transmitted on a frame-by-frame basis and formulate the problem as a QoE maximization problem with constraints on communication resources and long-term energy consumption. To tackle this problem, we propose a Lyapunov-guided LSTM-based DQN algorithm for bitrate adaptation. Experiments demonstrate the effectiveness of our proposed DQN algorithm, which outperforms baseline algorithms. In both the simulated system environment and the real system environment, we observed average QoE gains of 0.04 to 0.46 and 0.04 to 0.42 respectively. In the future, we will consider modelling video frame sizes more realistically, addressing the dual time-scale issue between high-level video frames and low-level wireless resources, and improving the QoE modelling to further enhance our research.
\color{black}

\bibliographystyle{IEEEtran}
\bibliography{reference}

\end{document}